# Scannerless non-line-of-sight three dimensional imaging with a 32×32 SPAD array

Chenfei Jin,* Meng Tang, Legeng Jia, Xiaorui Tian, Jie Yang, Kai Qiao and Siqi Zhang

School of Physics, Harbin Institute of Technology, Harbin 150001, China
*e-mail: <jinchenfei@hit.edu.cn>

Abstract

We develop a scannerless non-line-of-sight three dimensional imaging system based on a commercial 32×32 SPAD camera combined with a 70 ps pulsed laser. In our experiment, 1024 time histograms can be achieved synchronously in 3s with an average time resolution of about 165 ps. The result with filtered back projection shows a discernable reconstruction while the result using virtual wave field demonstrates a better quality similar to the ones created by earlier scanning imaging systems with single pixel SPAD. Comparatively, our system has large potential advantages in frame frequency, power requirements, compactness and robustness. The research results will pave a path for scannerless non-line-of-sight three dimensional imaging application.

In recent years, more and more research interests are focused on non-Line-of-Sight (NLoS) laser imaging with the ability of seeing around a corner, which is using reflections from glossy surfaces[1, 2] or diffuse surfaces [3] to extend the viewing into non-line-of-sight conditions. The NLoS imaging has many potential applications in robot vision, autonomous driving, remote sensing, security affairs and so on. At present, most NLoS imaging methods are based on precisely measuring time-of-flight (ToF) of the photons scattered from multiple surfaces along indirect paths, which need high detection sensitivity and time resolution of the sensor. Velten et al. [3, 4] firstly demonstrated the possibility to use ToF acquisition for NLoS imaging based on a streak camera with 2ps timing resolution. Nevertheless, the use of the streak camera is only limited in a laboratory environment due to the high cost, bulk and fragility. After that, other time-resolved devices also have been developed for NLoS imaging including intensified charge-coupled device (ICCD)[5,6], photonic mixer devices (PMDs) [7, 8], single-photon avalanche diode (SPAD)[9-13]. Especially in recent years, SPAD has been preferred for NLoS imaging due to its single photon detection capacity and picoseconds time resolution besides low cost and compactness. Many new methods such as anti-pinhole[14], LCT[15], F-K migration[16] and virtual wave field[17, 18] were all demonstrated by similar experimental setups of a single-pixel SPAD combined with scanned laser beam. However, Use of mechanical scanner could limit the imaging system to long acquisition time and high power consume. Recently, a 16 × 1 gated SPAD line array was specifically designed and experimentally demonstrated for NLoS imaging of a 40 × 60 cm hidden object by scanning over a grid of 150 × 150 positions[19]. Until now, existing commercial SPAD area array has only been used for tracking of NLoS targets[20], yet using it to acquire a reconstruction with a quality similar to the ones created with single pixel SPAD has rarely been demonstrated due to poor timing resolution, low number of pixels, small fill factor and lack of gating.

In this work, we developed an approach to scannerless NLoS 3D imaging. In our system, a commercial 32×32 SPAD area array was used as key sensor combined with a 70 ps pulsed laser. Without any scanner, the system finished ToF acquisition of the photons scattered from the hidden object in several seconds. The filtered back projection and virtual wave field were selected to process collected data for a high-resolution 3D reconstruction of an object.

As shown in Fig. 1, a NLoS experimental setup was realized consisting of a single-photon counting camera and a pulsed erbium fiber laser source. The image sensor consists of 1024 Si CMOS single-photon avalanche diodes arranged in a 32 × 32 area array. Single photon is sensed by the active area of each pixel with photon detection efficiency of 28% at 532nm. This active area is surrounded by the necessary electronics to bias and quench the SPAD, as well to time and count the detected photons but without gating, which leading to the only 2% fill factor of the sensor. The camera can be used in time-correlated single photon counting (TCSPC) mode in conjunction with a picosecond laser to perform time-resolved measurements. The laser is a versatile and flexible platform based on a Master Oscillator Fiber Amplifier (MOFA) concept with frequency conversion. The master oscillator generates infrared picosecond pulses at 1064 nm with variable repetition rates up to 80 MHz. The output of this seed laser is directly connected to a multi-stage fiber amplifier. The high pulse energies of the amplified 1064 nm infrared laser permit efficient wavelength conversions using second harmonic generation. In this way, it can generate 70ps pulses at 532nm with average optical power up to more than 300mW. It can work at 12 different internally selectable repetition rates between 31.25 kHz and 80 MHz and generates synchronous signals to trigger the image sensor. The ToF of a photon between its arrival and the arrival of the synchronous signal from the laser is measured and stored in the time histogram consisting of 1024 time bins with bin width of 55 ps.

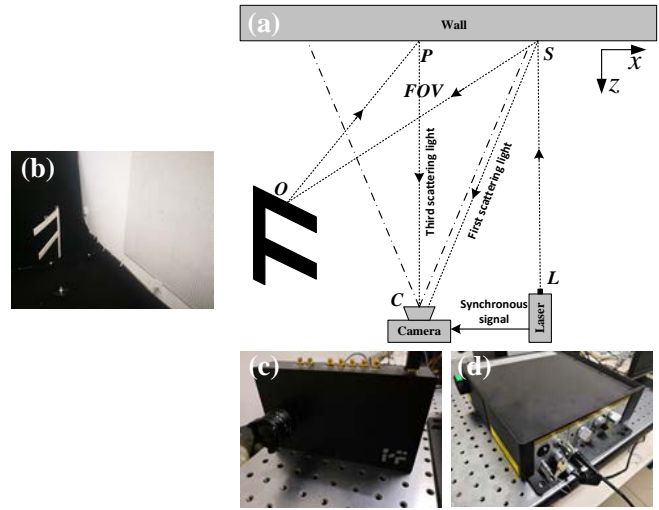

Fig. 1. (a)The scenario and the experimental setup including (b)the object, (c)the SPAD camera (d) the laser

A 1.5m×1.0m white foam board was fixed on the wall as an image screen. Both the camera and the laser were located at distance of 1.2 m from the screen. The scattered light from the image screen can be collected by a f/1.6-f/16 aperture lens with 1.8 mm-3.6 mm focal length. The field of view (FOV) of the camera was adjusted to cover a 1.0m×1.0m central region of the image screen. For avoiding the camera saturated, the laser beam with 0.5 mrad divergence angle was directed toward the screen and was firstly scattered at an illuminating point $S$ outside the field of view of the camera. A 0.5m×0.5m object 'F' was placed over a range of about 0.7-1m from the screen, outside the camera's direct field of view. The object surface was tilted by about 36° toward the normal vector of the wall. Several shielding screens and the floor around the object were covered by black light-absorbing cloth for removing the scatterings from non-interesting things. The secondly scattered light from the object 'F' was reflected thirdly on the screen, finally recorded by the camera.

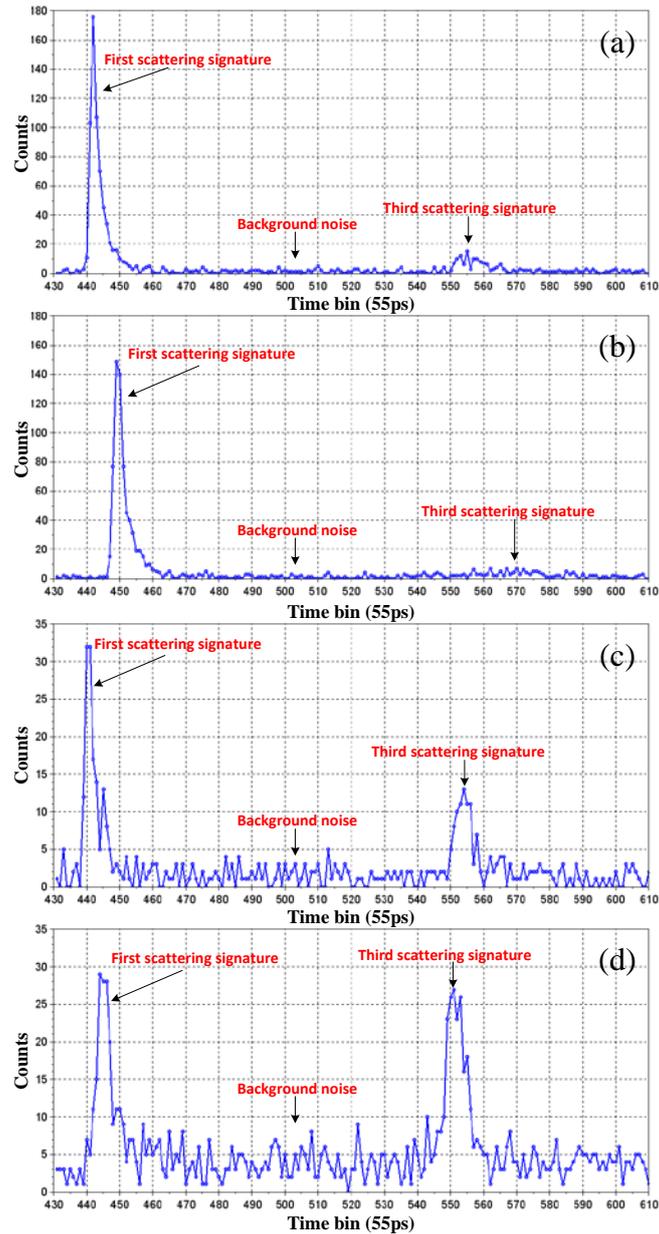

Fig. 2. Four typical time histograms for different pixels.

In the experiment, the laser was set up at 10MHz repetition rates with average optical power of about 40 mW and the camera was operated in 3s exposure time with the room lights off for obtaining the time histograms with high signal-to-noise ratio. The four typical time histograms for different pixels are shown in Fig. 2, respectively. The time histograms all consist of the third scattering signature encoding information of the object and the first scattering signature which is caused by direct reflection of the illuminating point. Although the illuminating point is not direct focused on the SPAD array, its direct reflection can go through a second bounce to arrive at each pixel almost simultaneously after entering into the camera. Fig. 2 shows the different level background noise which mainly comes from the ambient light and the dark count rate (DCR), the former is almost same and the latter is different for each sensor element. The DCR itself is a function of temperature and bias voltage, increasing of these parameters also increases the DCR. Under the default bias voltage, we measured the DCR of all pixels at room temperature as shown in Fig. 3(a). Correspondingly the statistical histogram of DCR is shown in Fig. 3(b), from which it can be calculated that about 80% of the pixels have the DCR lower than 100 counts/s and 90% of the pixels have the DCR lower than 1000 counts/s. In our experiment, the pixels with the DCR more than 1000 counts/s were thought as 'bad pixels', time histograms of which were replaced by interpolation of surrounding 'good pixels'.

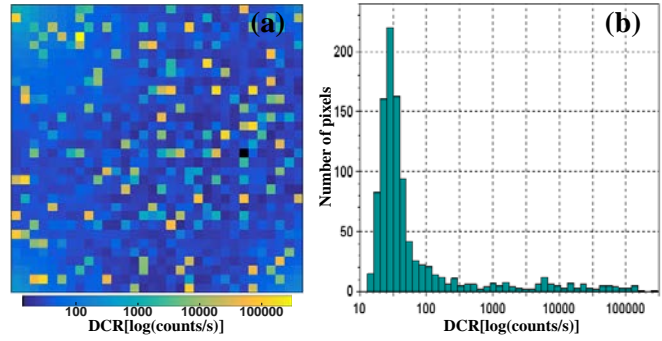

Fig. 3. (a) the spatial pattern and (b)the statistical histogram of DCR.

In addition, it can be obviously observed from Fig. 2 that peaks of the first scattering signatures are not always located at the same time, which mainly results from intrinsic time delay difference of instrument response function(IRF) for each pixel as shown in Fig. 4(a). Correspondingly the statistical histogram of peak locations of first scattering signatures is shown in Fig. 4(b), which indicates that the time delay differences among the pixels mainly cover a range of about 25 time bins (1375ps). In this case, about 41cm position error is produced when using time histograms of all pixels to reconstruct the object. For calibrating time delay difference, all time histograms were temporally aligned so that the first scattering signatures of these appeared at the same time.

As we know, Full Width at Half Maximum (FWHM) of IRF of a NLOS imaging system plays an important role on the resolution of reconstruction. The FWHM of our system mainly results from the pulse width $\tau_L$ (70 ps) of the laser and the time jitter $\tau_D$ (150 ps) of the SPAD which relates to the uncertainty in absolute timing of the stochastic avalanche process in response to the detection of a photon. Theoretical time resolution of 165.5 ps can be given with $\text{FWHM} = \sqrt{\tau_L^2 + \tau_D^2}$. The FWHM for each pixel was experimentally measured and shown in Fig. 5(a). Correspondingly the statistical histogram of FWHMs for all pixels is shown in Fig. 5(b), which demonstrates the FWHMs for all pixels cover a range of about 1-7 time bin. Average FWHM is about 3 time bins which corresponds to 165 ps time resolution, and it is about double ~ triple that of earlier NLoS scanning imaging systems with single pixel SPAD [15-17].

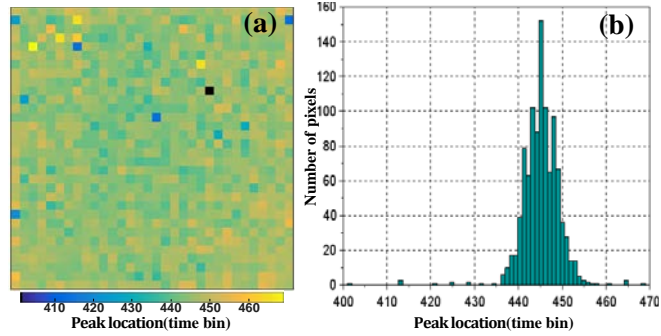

Fig. 4. (a) the spatial pattern and (b)the statistical histogram of peak location.

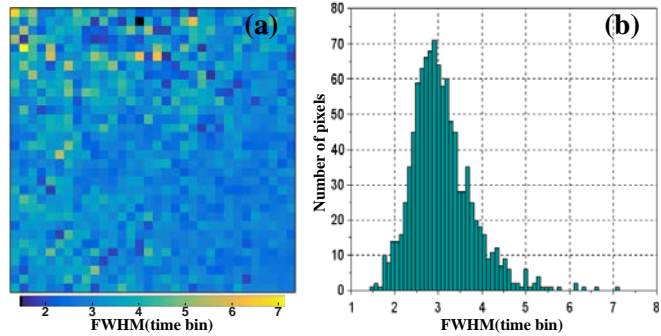

Fig. 5. (a) the spatial pattern and (b)the statistical histogram of FWHM.

The reconstruction algorithms adopted in the experiment contain the filtered back projection and virtual wave field which are all very suitable for our non-confocal NLoS imaging system based on the SPAD array. The filtered back projection is a popular standard method for NLoS imaging because it is free of assumption about the hidden scene geometry and has tractable computational time and memory requirements even for large-scale data[3-6,9-12,19-21]. In filtered back projection, the photons in each time bin of time histogram are successively projected into a voxelized space according to all possible flight paths of the photons to form a confidence map, which is further filtered for an exact reconstruction. The reconstruction results using filtered back-projection are shown in Fig. 6. The reconstruction region is limited in a cube of 1.2m×1.2m×1.2m divided into 120×120×120 voxels. The object 'F' with obvious distortions can be still distinguished from a clear background from front and side view, and top view shows a sharp edge with exact direction.

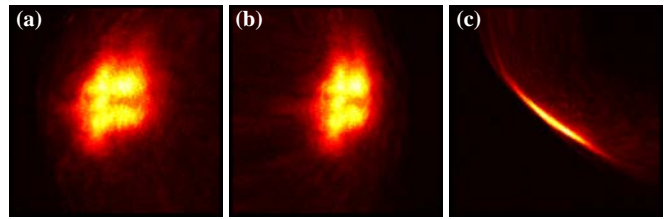

Fig. 6. Reconstructions using filtered back projection observed from (a) front view, (b) side view and

(c)top view.

Virtual wave field has recently been demonstrated with the scanning systems based on a single pixel SPAD[17, 18], and this method models the transient light transport in the hidden scene using a virtual wave equation so that the problem of NLoS imaging can be transformed into one of diffractive wave propagation. In our experiment, the selected parameters of virtual wave field had important effect on quality of the reconstruction as shown in Fig. 7.

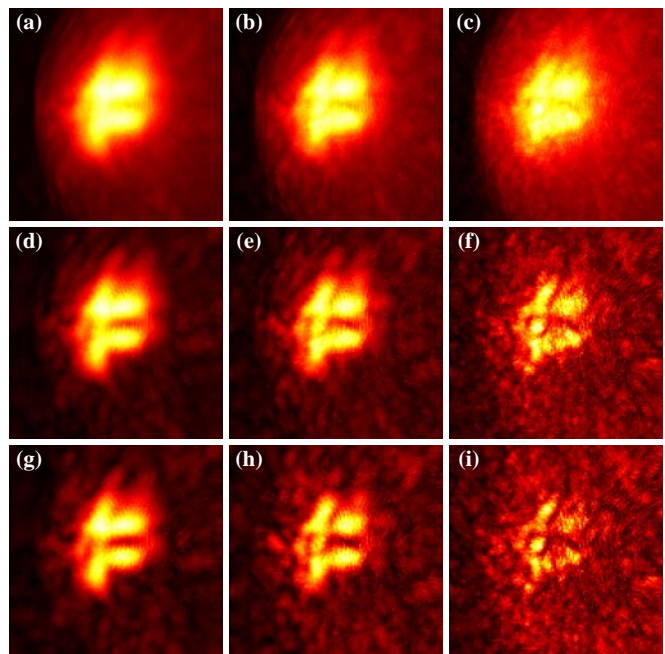

Fig. 7. Reconstructions using virtual wave field with (a)($\lambda$=10cm, $\sigma$=3), (b) ($\lambda$=8cm, $\sigma$=3), (c) ($\lambda$=6cm, $\sigma$=3), (d) ($\lambda$=10cm, $\sigma$=4), (e) ($\lambda$=8cm, $\sigma$=4), (f) ($\lambda$=6cm, $\sigma$=4), (g) ($\lambda$=10cm, $\sigma$=5), (h) ($\lambda$=8cm, $\sigma$=5)  (i) ($\lambda$=6cm, $\sigma$=5) observed  from  front view.

$\lambda$ is the wavelength of virtual wave field and $\sigma$ indicates the wavelet length of virtual wave field divided by the wavelength. It can be observed from Fig. 7 that the object is immersed in a speckle-like background. As the wavelength is reduced, the resolution of the object's details is improved but the contrast between the object and the speckles gets worse. In this case, more shape distortions will be produced on the object. As the wavelet length is increased, the discriminability of the object's edges is improved but the contrast between the object and the speckles gets also worse. In this case, some edges are apt to disappear from the object. After repeatedly trying, the optimum parameters combination was found for optimal reconstruction using virtual wave field as is shown in Fig. 8. The object 'F' with few distortions and deficiencies can be clearly distinguished from the speckles from the front and side view, and top view also shows an obvious edge with exact direction. Compared to the reconstruction results of filtered back projection, those of virtual wave field demonstrates more exact shape feature with higher resolution, and it even has been up to a similar quality to the ones created by earlier scanning imaging systems with single pixel SPAD.

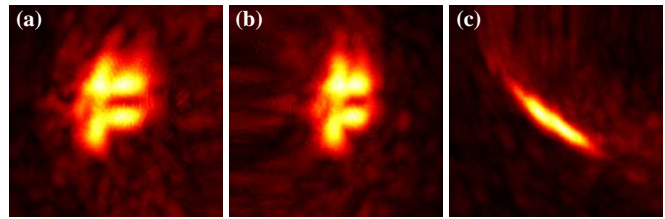

Fig. 8. Reconstructions using virtual wave field with ($\lambda$=9.6cm, $\sigma$=4.7) observed from (a) front view, (b) side view and (c) top view.

In summary, we developed a scannerless NLoS 3D imaging prototype. In our system, a commercial 32×32 SPAD camera is used as the key sensor combined with a ps pulsed laser. Without any scanner, the camera finished ToF acquisition of the photons scattered from the hidden object in 3 seconds. The difference of dark count, time delay and FWHM among all pixels were measured and analyzed according to the collected data. In reconstruction process, virtual wave field was demonstrated with better reconstruction quality than that of filtered back projection. Compared to earlier scanning imaging systems with single pixel SPAD, our system has large potential advantages in frame frequency, power requirements, compactness and robustness. Our research will pave a path for scannerless NLoS imaging application.

## References

1. O. Steinvall, M. Elmqvist, and H. Larsson, "See around the corner using active imaging," Proc. SPIE **8186**, 818605 (2011).

2. V. Molebny, and O. Steinvall, "Multi-dimensional laser radars," Proc. SPIE **9080**, 908002(2014).

3. A.Velten, T. Willwacher, O. Gupta, A. Veeraraghavan, M. G. Bawendi, and R. Raskar, "Recovering Three-dimensional Shape Around a Corner using Ultrafast Time-of-Flight Imaging," Nat. Commun. **3**, 745(2012).

4. O. Gupta, T. Willwacher, A. Velten, A. Veeraraghavan, and R. Raskar, "Reconstruction of hidden 3D shapes using diffuse reflections, " Opt. Express **20**(17), 19096-19108(2012).

5. M. Laurenzis and A. Velten, "Non-line-of-sight active imaging of scattered photons," Proc. SPIE **8897**, 889706 (2013).

6. M. Laurenzis and A. Velten, "Non-line-of-sight laser gated viewing of scattered photons, " Opt. Eng. **53**(2), 023102 (2014).

7. F. Heide, M. B. Hullin, J. Gregson and W. Heidrich, "Low-budget Transient Imaging using Photonic Mixer Devices," ACM Transactions on Graphics **32**(4), 45(2013).

8. F. Heide, L. Xiao, W. Heidrich, and M. B. Hullin, "Diffuse Mirrors: 3D Reconstruction from Diffuse Indirect Illumination Using Inexpensive Time-of-Flight Sensors, " in *Proceedings of IEEE Conference on Computer Vision and Pattern Recognition* (IEEE, 2014) 3222-3229.

9. M. Buttafava, J. Zeman, A. Tosi, K. Eliceiri, and A. Velten, "Non-line-of-sight imaging using a time-gated single photon avalanche diode, " Opt. Express **23**(16), 20997-21011(2015).

10. S. Chan, R. E. Warburton, G. Gariepy, J. Leach, and D. Faccio, "Non-line-of-sight tracking of people at long range," Opt. Express **25**(9), 10109-10117(2017).

11. C. Jin, J. Xie, S. Zhang, Z. Zhang, and Y. Zhao, "Reconstruction of multiple non-line-of-sight objects using back projection based on ellipsoid mode decomposition, " Opt. Express **26**(16), 20089-20101 (2018).

12. C. Jin, J. Xie, S. Zhang, Z. Zhang, and Y. Zhao, "Richardson–Lucy deconvolution of time histograms for high-resolution non-line-of-sight imaging based on a back-projection method, " Opt. Lett.**43**(23), 5885-5888(2018).

13. G. Musarra, A. Lyons, E. Conca, Y. Altmann, F. Villa, F. Zappa, M.J. Padgett, and D. Faccio, "Non-Line-of-Sight Three-Dimensional Imaging with a Single-Pixel Camera," Phys. Rev. Appl. **12**(1), 011002 (2019).

14. F. Xu, G. Shulkind, C. Thrampoulidis, J. H. Shapiro. A. Torralba, F.N.C. Wong, and G.W. Wornell. "Revealing hidden scenes by photon-efficient occlusion-based opportunistic active imaging," Opt. Express **26**(8), 9945-9962(2018).

15. M. O'Toole, D. B. Lindell, and G. Wetzstein, "Confocal non-line-of-sight imaging based on the light-cone transform," NATURE **0**, 1-4 (2018)

16. D. B. Lindel, G. Wetzstein, M. O'Toole, "Wave-based non-line-of-sight imaging using fast f–k migration," ACM Trans. Graph. **38**, 116 (2019).

17. X. Liu, L. Guillén, M. L. Manna, J. H. Nam, S. A. Reza, T. H. Le, A. Jarabo, D. Gutierrez, and A. Velten, "Non-line-of-sight imaging using phasor- field virtual wave optics, " Nature **572**, 620–623 (2019).

18. M. L. Manna, J. Nam, S. A. Reza, and A. Velten, "Non-line-of-sight-imaging using dynamic relay surfaces," Opt. Express **28**(4), 5331-5339(2020).

19. M. Renna, J. H. Nam, M. Buttafava , F. Villa, A. Velten and A. Tosi, "Fast-Gated 16×1 SPAD Array for Non-Line-of-Sight Imaging Applications," Instruments **4**(2), 14(2020).

20. G. Gariepy, F. Tonolini, R. Henderson, J. Leach, and D. Faccio, "Detection and tracking of moving objects hidden from view," Nat. Photonics **10**, 23-27(2016).

21. V. Arellano, D. Gutierrez, and A. Jarabo, "Fast back-projection for non-line of sight reconstruction," Opt. Express **25**(10), 11574-11583(2017).